\def\rn{}
\def\nn#1 #2{#2. #1}				         % Name with 1 initial
\def\nnn#1 #2 #3{#2. #3. #1}			% Name with 2 initials
\def\nnnn#1 #2 #3 #4{#2. #3. #4 #1}		% Name with 3 initials
\def\nnnnn#1 #2 #3 #4 #5{#2. #3. #4 #5. #1}	% Name with 4 initials
\def\multiand{, and\hbox{ }}				
\def\rf#1;#2;#3;#4;#5 {{\frenchspacing\par\rn#1, #3 {\bf #4}, #5 (#2). \par}}
\def\rg#1;#2;#3;#4;#5;#6 {{\frenchspacing\par\rn#1, #3 {\bf #4}, #5 (#2). \par}}
\def\rfbook#1;#2;#3;#4;#5 {{\frenchspacing\par\rn#1, {\it #3} (#5, #4, #2).\par}}
\def\rfprep#1;#2;#3 {{\par\frenchspacing\rn#1, #3 (#2).\par}}
\def\preskip{\vskip-0.0cm}

\def\beq#1{\begin{equation}\label{#1}}
\def\eeq{\end{equation}}
\def\eq#1{equation~(\ref{#1})}
\def\Eq#1{Equation~(\ref{#1})}

\def\beqa#1{\begin{eqnarray}\label{#1}}
\def\eeqa{\end{eqnarray}}
\def\bt{\begin{tabbing}}
\def\et{\end{tabbing}}

\def\Sec#1{Section~\ref{#1}}

\def\Sub;tion#1{{\vskip-0.8cm}\hbox{\,}\subsection{#1}}

\def\bfig{\begin{figure}[h] \centerline{\hbox{}}\vfill}
\def\efig{\end{figure}\vfill\newpage}

\def\spose#1{\hbox to 0pt{#1\hss}}
\def\simlt{\mathrel{\spose{\lower 3pt\hbox{$\mathchar"218$}} \raise 2.0pt\hbox{$\mathchar"13C$}}}
\def\simgt{\mathrel{\spose{\lower 3pt\hbox{$\mathchar"218$}} \raise 2.0pt\hbox{$\mathchar"13E$}}}
\def\simpropto{\mathrel{\spose{\lower 3pt\hbox{$\mathchar"218$}} \raise 2.0pt\hbox{$\propto$}}}

\def\etal{{\frenchspacing\it et al.}}

\def\eg{{\frenchspacing\it e.g.}}
\def\ie{{\frenchspacing\it i.e.}}

\def\uK{\mu{\rm K}}
\def\l{\ell}

\def\alm{a_{\l m}}
\def\atlm{\tilde{a}_{\l m}}
\def\lth{\l^{\rm th}}

\def\nh{\widehat{\bf n}}
\def\A{{\bf A}}
\def\B{{\bf B}}
\def\D{{\bf D}}
\def\L{{\bf L}}
\def\U{{\bf U}}
\def\R{{\bf R}}
\def\Zero{{\bf 0}}
\def\dmin{{d_{\rm min}}}
\def\expec#1{\langle#1\rangle}

\def\etal{{\frenchspacing\it et al.}}
\def\ie{{\frenchspacing\it i.e.}}
\def\eg{{\frenchspacing\it e.g.}}

\documentclass[twocolumn,amsmath,nofootinbib]{revtex4}
\usepackage{url,epsf}
\begin{document}
%\twocolumn[\hsize\textwidth\columnwidth\hsize\csname@twocolumnfalse\endcsname

\title{The significance of the largest scale CMB fluctuations in WMAP}
\author   {Ang\'elica de Oliveira-Costa$^1$, Max Tegmark$^1$, Matias Zaldarriaga$^2$ \& Andrew Hamilton$^3$}
\address{$^1$Dept. of Physics \& Astronomy, University of Pennsylvania, Philadelphia, PA 19104, USA}
\address{angelica@higgs.hep.upenn.edu}
\address{$^2$Dept. of Physics, Harvard University, Cambridge, MA 02138, USA}
\address{$^3$JILA \& Dept. of Astrop. \& Planetary Sciences, U. Colorado, Boulder, CO 80309 , USA}
\date       {Submitted to Phys. Rev. D. July 15 2003, revised Oct. 14 2003 - still no referee report.}

%%%%%%%%%%%%%%%%%%%%%%%%%%%%%%%%%%%%%%%%%%%%%%%%%%
%%%%%%%%%%%%%%%%%%%%%%%%%%%%%%%%%%%%%%%%%%%%%%%%%%

\begin{abstract}
We investigate anomalies reported in the Cosmic Microwave Background 
maps from the Wilkinson Microwave Anisotropy Probe (WMAP) satellite on 
very large angular scales and discuss possible interpretations. Three 
independent anomalies involve the quadrupole and octopole:
{\it 1)} The cosmic quadrupole on its own is anomalous at the 1-in-20 level by 
        being low (the cut-sky quadrupole measured by the WMAP team is more 
        strikingly low, apparently due to a coincidence in the orientation of our 
        Galaxy of no cosmological significance);
{\it 2)} The cosmic octopole on its own is anomalous at the 1-in-20 level by 
         being very planar; 
{\it 3)} The alignment between the quadrupole and octopole is anomalous 
        at the 1-in-60 level.
Although the {\it a priori} chance of all three occurring is 1 in 24000, the 
multitude of  alternative anomalies one could have looked for dilutes the 
significance of such {\it a posteriori} statistics.
The simplest small universe model where the universe has toroidal topology 
with one small dimension of order half the horizon scale, in the direction towards 
Virgo, could explain the three items above. However, we rule this model out using 
two topological tests: the $S$-statistic and the matched circle test.
%In particular, our results rule out the recently proposed 
%dodecahedron model of Luminet, Weeks, Riazuelo, Lehoucq \& Uzan.
\bigskip
\end{abstract}

\keywords{cosmic microwave background  -- topology}
%%-- radiation mechanisms: thermal and non-thermal
%%-- methods: data analysis}
\pacs{98.80.Es}
%]%%% End front material
  
%%%%%%%%%%%%%%%%%%%%%%%%%%%%%%%%%%%%%%%%%%%%%%%%%%
%%%%%%%%%%%%%%%%%%%%%%%%%%%%%%%%%%%%%%%%%%%%%%%%%%

\maketitle

\section{Introduction}

It is common in science that when measurements are improved, old
questions are answered and new ones are raised. In this sense, 
history has repeated itself with the spectacular new Cosmic Microwave 
Background (CMB) results from the Wilkinson Microwave Anisotropy Probe
(WMAP) \cite{bennett03a}. 
The WMAP power spectrum results \cite{hinshaw03} are in near-perfect
agreement with the cosmological concordance model in {\it vogue} 
(see, \eg, \cite{wang01,efstathiou01}), but have sent cosmologists 
scrambling to figure out what to make of the detection of a high reionization 
optical depth and hints of a running spectral index \cite{spergel03,verde03}.
The ``near-perfect'' caveat above refers to the surprisingly low amplitude
of the observed CMB quadrupole, reflecting a lack of large-angle
correlations significant at about the 99.9\% level \cite{hinshaw03,spergel03}, 
{\it c.f.} \cite{enrique03,efstathiou03b}.
The low quadrupole amplitude seen by WMAP was also observed by 
COBE/DMR (see, \eg, \cite{smoot92}, \cite{hinshaw96} and references 
therein), but only now, with the large WMAP frequency coverage and 
the low detector noise, has it become  possible to show that this low 
CMB quadrupole is not significantly affected by Galactic emission 
\cite{bennett03b,tegmark03,enrique03}.  Compounding the puzzle, 
anomalies related to the WMAP octopole  were reported in \cite{tegmark03}.

\begin{figure*}[tb]
\center{{\hglue-0.5cm\epsfxsize=19.0cm\epsffile{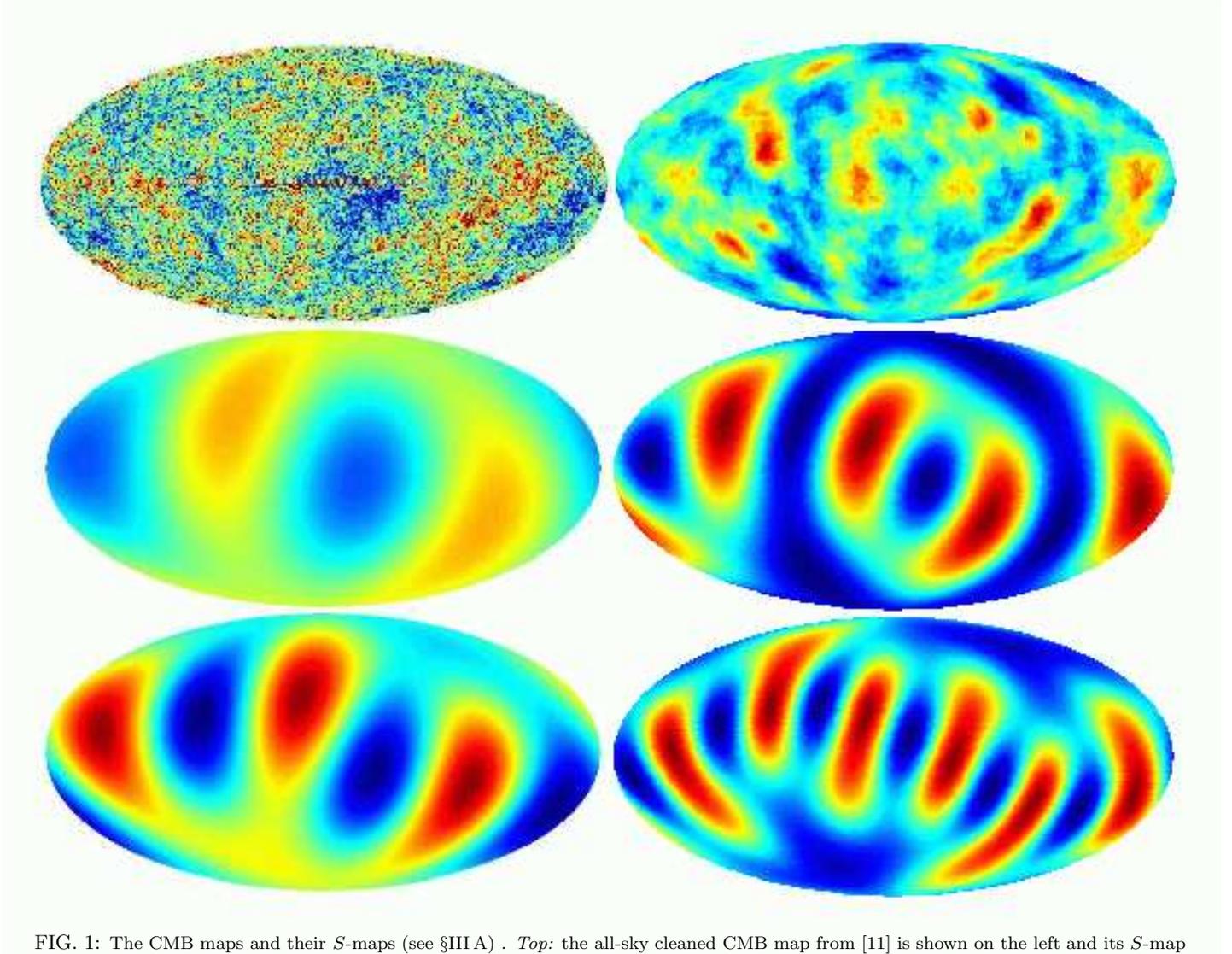}}}
%\begin{figure}[tb]
%\preskip
%\centerline{\epsfxsize=9.2cm\epsffile{mybabamaps.eps}}
\vskip-0.8cm
\caption{\label{fig1}\footnotesize%
	    The CMB maps and their $S$-maps (see \S\protect\ref{BabaSec}) . 
              {\it Top:} the all-sky cleaned CMB map from \protect\cite{tegmark03} 
	     is shown on the left and its $S$-map on the right. 
	     {\it Middle:} the quadrupole map (left) and its $S$-map (right). 
	     {\it Bottom:} the octopole map (left) and its $S$-map (right).
	     Note that all three $S$-maps show dark spots in the supposed direction
	     of suppression of its original maps, around ``two o'clock''.
              }
%\end{figure}
\end{figure*}

During years, the suppression of large-scale power (sometimes referred in 
the literature as a {\it cutoff}), have been used by many authors as evidence 
for a ``small universe" with non-standard topology (see, \eg,
 \cite{sokolov93,fang93,stevens93,doc95,graca02}).  If so, the conventional 
explanation is that such suppression could be caused by the fact that we live 
in a Universe with compact topology, in which power on scales exceeding the 
fundamental cell size is suppressed -- see \cite{janna01} and references 
therein for a review of this subject.
If the low multipole behavior is caused by our universe having $T^1$ 
\cite{doc96} compact topology with one dimension small relative to the 
horizon scale, then the large scale power would be suppressed in this 
particular spatial direction as illustrated in Figure~\ref{fig1}. 
Another possibility is hyperspherical topology corresponding to a closed 
universe \cite{uzan03b,efstathiou03}.

Another class of explanations that have been proposed involve modifying
the inflationary picture to introduce a cutoff in the primordial power spectrum 
\cite{bridle03,uzan03a,contaldi03,cline03,feng03}, perhaps linked to the 
spatial curvature scale \cite{efstathiou03} or string physics \cite{bastero03}.
There is also a class of models that explain the lower multipole values by 
invoking the existence of fluctuations in a quintessence-like scalar field 
which can introduce features on scales comparable to the present day 
Hubble radius \cite{dedeo03}. The difficulty here is to cancel the large 
integrated Sachs-Wolfe anisotropy associated with a high vacuum energy 
density. 

The goal of this paper is to go back to the WMAP data and investigate 
these large-scale anomalies in greater detail, to clarify what anomalies
(if any) are statistically significant and to look for other signatures of 
non-standard topologies. 
The rest of this paper is organized as follows. In \Sec{AnomalySec}, we 
compute the significance of the low quadrupole, the planar octopole and 
their alignment. In \Sec{BabaSec}, we simulate a grid of small-universe 
models to quantify whether they fit the data by employing a statistic that 
searches for topology-induced symmetries in the CMB sky \cite{doc96}.
In \Sec{CircleSec}, we search for matched circles in the CMB sky, an idea 
proposed by \cite{css98}, ruling out the simplest class of small-universe 
models. Finally, we present our conclusions in \Sec{ConclusionSec}.

\clearpage

\begin{table}[tb]
\caption{\footnotesize
	CMB quadrupole and octopole power.}\label{tab1}
\begin{center}
{\footnotesize
\begin{tabular}{lccr}
\hline\hline
Measurement &$\delta T_2^2$ $[\mu$K$^2]$	&p-value	&$\delta T_3^2$ $[\mu$K$^2]$\\
\hline		
Spergel {\etal} model	&869.7 	&		         & 855.6\\
Hinshaw {\etal} cut sky      &123.4	&0.7\%		& 611.8\\
ILC map all sky 	         &195.1	&4.8\%		&1053.4\\
%Cleaned map all sky  	&201.6	&5.1\%		& 866.1\\
Tegmark {\etal}  	         &201.6	&5.1\%		& 866.1\\
Cosmic quadrupole	 	&194.2	&4.7\%		&\\
Dynamic quadrupole	&  3.6	&		         &\\
\hline
\hline		
\end{tabular}}
\end{center}
\end{table}

\section{How significant are the ``anomalies'' in the lower multipoles?}\label{AnomalySec}

\subsection{The low quadrupole}\label{QuadrupoleSec}

The surprisingly small CMB quadrupole has intrigued the cosmology 
community ever since it was first observed by COBE/DMR \cite{smoot92}. 
However, it was not until the precision measurements of WMAP \cite{bennett03b} 
that it became clear that the anomaly could not be blamed on Galactic foreground 
contamination.  
Simulations by \cite{spergel03,verde03} have demonstrated that within the 
context of the standard inflationary $\Lambda$CDM concordance model,
the low large-scale power observed is sufficiently unlikely to warrant serious 
concern. The WMAP team argued that the low quadrupole requires a fluke 
at the one in 143 level and the low overall large-scale power (mainly 
quadrupole and octopole) require a one in 666 fluke \cite{spergel03}. 

In \cite{tegmark03}, it was argued that this deficit of large-scale power was 
of lower statistical significance than previously claimed, and this conclusion
has been strengthened by \cite{enrique03,efstathiou03b}. 
In particular,  \cite{efstathiou03b} gives a thorough discussion of this statistical 
issue in the context of both frequentist and Bayesean analysis, concluding 
that the WMAP results are in fact consistent with the concordance  
$\Lambda$CDM model. Here, we will further elaborate the foreground-related 
point raised in \cite{tegmark03}.
The WMAP team measured the angular power spectrum $C_\l$ using only the 
part of the sky outside of their Galactic cut, and \cite{tegmark03} found that this 
cut, seemingly coincidentally, eliminated the strongest hot/cold spots of the 
quadrupole and octopole.
To correctly interpret the low {\it a priori} likelihood of the WMAP measurements, 
we need to factor it as a product of two different probabilities, corresponding to
the following two questions:
\begin{enumerate}
	{\it 
	\item How unlikely is the all-sky power spectrum given a cosmological 
                   model like $\Lambda$CDM?
	\item Given the all-sky CMB map, how unlikely is it that the Galactic cut 
		will eliminate such a large fraction of the quadrupole and octopole 
		power?
	}
\end{enumerate}
The key point is that the location of the Galactic cut is determined by the 
orientation of the Milky Way, \ie, by fluctuations that have nothing to 
do with the CMB quadrupole and octopole. This means that although an 
unlikely coincidence associated with the second question may be disturbing,
it is of no cosmological significance and should be ignored when testing 
cosmological models.

\begin{table}[tb]
\caption{\footnotesize
	Observed CMB quadrupole and octopole coefficients
	in the foreground-cleaned WMAP map \cite{tegmark03}
	calculated in Galactic coordinates in units of $\uK$.}\label{tab2}
\begin{center}   
{\footnotesize
\begin{tabular}{lccr}
\hline\hline
$\l$	&$m$ &Re$\{\alm\}$	 &Im$\{\alm\}$\\
\hline
2	&0	& 10.73     &    0.00\\
2	&1	&  -5.87     &    4.26\\
2	&2	&-13.71     &-15.15\\
3	&0	&  -6.52     &    0.00\\
3	&1	&  -9.08     &    0.68\\
3	&2	& 21.57     &    1.73\\
3	&3	&-13.64     & 28.79\\
\hline
\hline
\end{tabular}}
\end{center}
\end{table}

Table~\ref{tab1} summarizes measurements of the quadrupole (column 2) and 
octopole (column 4) power reported in \cite{spergel03,hinshaw03,tegmark03}. 
% The 4th row in this table corresponds to the numbers shown in Table~\ref{tab2}, 
% with the relation being
Table~\ref{tab2} gives a breakdown of the coefficients $\alm$ which produce the 
quadrupole and octopole power given in row 4 of Table~\ref{tab1}, according to

$$\delta T_\l^2=\l(\l+1)C_\l/2\pi,$$
\noindent with \\
$$C_\l=(2\l+1)^{-1}\sum_{\l=-m}^m |\alm|^2.$$\\
The third column of Table~\ref{tab1} shows the probability of the quadrupole 
in our Hubble volume being as low as observed if the best-fit WMAP team 
model from \cite{spergel03,verde03} is correct. For the all-sky case 
where $\delta T^2_2$ has a $\chi^2$-distribution with 5 degrees of 
freedom, the probability tabulated is simply 

$$1 - \gamma[5/2,(5/2) T^2_2/855.6\uK^2]/\Gamma(5/2),$$\\
where $\gamma$ and $\Gamma$ are the incomplete and complete 
Gamma functions, respectively.
According to Question~{\it 1} , the low cosmic quadrupole requires 
about a one in 20 coincidence, and the cosmic octopole is not 
low at all, actually exceeding the theoretical prediction.
The one in 666 coincidence from \cite{spergel03} thus factors roughly 
into a one in 20 cosmic coincidence (Question~{\it 1}) and a one in 33 
Galactic orientation coincidence (Question~{\it 2}).

Foreground modeling remains the dominant uncertainty underlying 
these arguments. Although the detailed tests reported in \cite{tegmark03} 
suggested that the quadrupole and octopole contributions from the 
inner Galactic plane where unimportant (after multifrequency foreground 
subtraction), this issue deserves further study. Moreover, exotic foreground 
processes that are not localized to the Galactic plane, say SZ-emission
from the Galactic halo or the local super-cluster, would not have been 
detected in these tests. If a substantial fraction of the all-sky power 
reported in Table~\ref{tab2} turns out to be attributable to contamination, this 
will of course exacerbate the problem for the $\Lambda$CDMs 
cosmological model. 

\subsection{The quadrupole-octopole alignment}\label{AlignmentSec}

As seen in Figure~\ref{fig1}, the quadrupole of the foreground-cleaned 
WMAP map \cite{tegmark03} (at middle left in the figure) and 
octopole (at bottom left) both appear rather planar, with most of their 
hot and cold spots centered on a single plane in the sky\footnote{
	Note that this is planarity is somewhat 
	obscured by the Aitoff projection.}.
Moreover, the two planes appear roughly aligned. 
{\it  Can we quantify the statistical significance of this alignment?}

A simple way to quantify a preferred axis for arbitrary 
multipoles is to think of the CMB map as a wave function

$$ {\delta T\over T}(\nh) \equiv \psi(\nh)$$\\ 
and find the axis $\nh$ around which the angular momentum dispersion
\beq{Leq}
	\expec{\psi|(\nh\cdot\L)^2|\psi} = \sum_{m} m^2 |a_{\l m}(\nh)|^2
\eeq
is maximized. Here $a_{\l m}(\nh)$ denotes the spherical harmonic 
coefficients of the CMB map in a rotated coordinate system with its 
$z$-axis in the the $\nh$-direction. In practice, we perform the maximization 
by evaluating \eq{Leq} at all the unit vectors $\nh$ corresponding to 
$3{,}145{,}728$ HEALPix\footnote{
	The HEALPix package is available from
	\protect\url{http://www.eso.org/science/healpix/}.}
pixel centers at resolution $nside$=512, which pinpoints the maximum 
to within half the pixel spacing, about 0.06 degrees. Table~\ref{tab2} lists the 
coefficients $\alm$ for $\l=2,3$ in Galactic coordinates. We then use 
Wigner's formula (see Appendix A) to rotate these coefficients into 
other coordinate systems, which is much faster than repeatedly rotating 
the entire sky map. We find the preferred axes $\nh_2$ and $\nh_3$ for 
the quadrupole and octopole, respectively, to be 
\beqa{axisEq}
	\nh_2&=& (-0.1145, -0.5265, 0.8424), \nonumber\\
	\nh_3&=& (-0.2578, -0.4207, 0.8698), 
%\nh_2&=& (-0.114525452 -0.526464939 0.842447996)
%\nh_3&=& (-0.257810831 -0.420708865 0.869791687).
\eeqa
\ie, both roughly in the direction of $(l,b)\sim (-110^\circ,60^\circ)$ in Virgo.
% Quadruople:  (l,b)=(257.727264,57.3995247)
% Octopole:       (l,b)=(238.499985,60.4344406)
% Average:        (l,b)=(248,59)=(-112,59)

Under the null hypothesis that the CMB is an isotropic random field, the 
quadrupole and octopole are statistically independent, with the unit 
vectors $\nh_2$ and $\nh_3$ being independently drawn from a 
distribution where all directions are equally likely. This means that 
the dot product $\nh_2\cdot\nh_3$ is a uniformly distributed random 
variable on the interval $[-1,1]$\footnote{
	This is most easily seen as follows. In a coordinate system where 
	$\nh_2=(0,0,1)$ and 
	$\nh_3=(\sin\theta\cos\varphi,\sin\theta\sin\varphi,\cos\theta)$,
	the dot product is $\nh_2\cdot\nh_3=\cos\theta$ which has a 
	uniform distribution, since the area element 
	$d\Omega=\sin\theta d\theta d\varphi \propto d\cos\theta 
	                  = d(\nh_2\cdot\nh_3)$.
	}.
\Eq{Leq} does not distinguish between $\nh$ and $-\nh$ (we find a 
preferred axis, not a preferred direction), so let us quantify the alignment 
by studying the quantity $|\nh_2\cdot\nh_3|$ which has a uniform 
distribution on the unit interval $[0,1]$. \Eq{axisEq} gives 
	$|\nh_2\cdot\nh_3|\approx  0.9838$,      % 0.9837686419
corresponding to a separation of $10.3^\circ$.
An alignment this good happens by chance only once in 
	$1/(1-0.9838)\approx 62.$ 
In other words, the probability that a random axis falls inside a circle 
of radius $10.3^{\circ}$ around the quadrupole axis is simply the area 
of this circle over the area of the half-sphere, $1/62$.

\begin{table}[tb]
\caption{\footnotesize
	Real-valued quadrupole and octopole coefficients 
	$\atlm$ (defined in Appendix A) in Galactic coordinates (left) and 
	rotated into their preferred frame of \eq{axisEq} (right), in units of 
	$\uK$. The observed $\delta T_\l$ for $\l=2,3$ are the rms of 
	these numbers times $[\l(\l+1)/(2\pi)]^{1/2}.$}\label{tab3}
\begin{center}   
{\footnotesize
\begin{tabular}{lccr}
\hline\hline
$\l$	&$m$ &Galactic &Rotated\\
\hline
2	&-2	&-21.43    & 13.32\\
2	&-1	&  6.03      & -0.40\\
2	&0	& 10.73     &  6.72\\
2	&1	& -8.30      & -0.40\\
2	&2	&-19.39     & 28.86\\
\hline
3	&-3	& 40.71     & 50.58\\
3	&-2	&  2.45      & -1.67\\
3	&-1	&  0.96      & -0.50\\
3	&0	& -6.52      &-13.60\\
3	&1	&-12.84     & -0.27\\
3	&2	& 30.50     &  0.71\\
3	&3	&-19.29     &-20.68\\
\hline\hline
\end{tabular}}
\end{center}  
\end{table}

Although \cite{tegmark03} argued that residual foreground contamination 
is not likely to dominate the quadrupole or octopole, it is important to keep 
in mind the possibility that some form of residual foreground contamination 
might nonetheless contribute to the alignment of the two. Note that the 
plane orthogonal to the preferred axes given by  \eq{axisEq} is about 
$\cos^{-1} 0.85\approx 30^\circ$ away from the Galactic plane, which 
is not particularly significant (requiring only a 1 in 6 coincidence).

\subsection{The planar octopole}\label{OctopoleSec}

We just discussed the statistical significance of two anomalies: an intrinsic 
property of the quadrupole (being low) and the alignment between the 
quadrupole and octopole. Figure~\ref{fig1}, however, suggests 
the presence of a third anomaly intrinsic to the octopole: being unusually 
planar.
In contrast, the hexadecapole and higher multipoles seem to exhibit a 
more generic behavior, as we expect in an isotropic random field, with no 
obvious preferred axis.  
{\it How unlikely is for an octopole to be so planar by chance?} 

According to the first statistical test we performed, using the symmetry 
statistic that we will describe in detail in the next section, this requires 
a one in 20 coincidence in the context of a Gaussian random field.
Another obvious test involves the angular momentum used in \eq{Leq}. 
We define a test statistic $t$ that measures 
the maximal percentage of the octopole power that can be attributed to 
$|m|=3$, \ie, 
\beq{tEq}
	t\equiv \max_{\nh} {|a_{3-3}(\nh)|^2+|a_{33}(\nh)|^2\over
	\sum_{m=-3}^3|a_{3m}(\nh)|^2}, 
\eeq
and find that for the cleaned WMAP map \cite{tegmark03} $t$=94\%.
Performing a large number of Monte Carlo simulations corresponding to 
an isotropic Gaussian random field (where the seven real-valued
coefficients $\tilde{a}_{3m}$ are simply independent Gaussian random 
variables with zero mean and identical variance), we obtain values larger 
than this about 7\% of the time.
Redefining $t$ as the ratio of the angular momentum dispersion to the 
total power, \ie, 
	$t\equiv \expec{\psi|(\nh\cdot\L)^2|\psi}/\expec{\psi|\psi}$, 
gives a similar significance level. In both of these tests, the planar nature 
of the octopole is reflected by the dominant contribution from $|m|=3$ to 
the octopole rms in the last column of Table~\ref{tab3}. In contrast, the 
quadrupole is not significantly planar according to these same statistical 
tests.

% An alternative is diagonalizing the quadrupole, but that of course only works for $\l=2$.
% (talk about how a quadrupole is a quadratic form)
% But we just want to do ONE test, the first that comes to mind, to avoid 
% data-mining bias.

\section{Topological signatures} 

In the previous section, we investigated the anomalies reported in the 
maps from the WMAP satellite on very large angular scales. 
Although the {\it a priori} chance of all these three anomalies occurring is 
1 in 24000, the multitude of alternative anomalies one could have looked 
for dilutes the significance of such {\it a posteriori} statistics.
It is important to point out, however, that in 
the simplest small Universe model 
where the Universe has toroidal topology with one small dimension smaller 
than the horizon scale, one would expect all three of the  
anomalies discussed above. This means that in the context of 
constraining cosmic topology, these three effects are not merely 
three arbitrary ones among many.

Two completely separate observable signatures of small universes 
have been described in the literature. 
The first, known as the $S$-statistic  \cite{baba93,doc96}, assumes that 
modifications of the CMB power spectrum are caused by the fact that 
only certain fluctuation modes are allowed by the boundary conditions, 
just as the fundamental tone and its overtones give information about 
the geometry inside a flute. 
The second, known as circles-in-the-sky \cite{css98},  is a purely 
geometric effect in which a given space-time point may be observable 
by us in more than one direction.
 
These two signatures are both independent and complementary. The 
information about the first comes from the largest angular scales in the 
CMB, while information about the second comes from smaller-scale
CMB patterns, as well as distant objects such as quasars that could 
potentially be multiply imaged (see, \eg, \cite{fagundes87,weatherley03}). 
Detection of a signature of the second kind could provide smoking gun 
evidence of a small universe, being independent of any assumptions 
about cosmological perturbation theory and potentially providing high 
statistical significance.

\subsection{Searching for symmetries in the sky}\label{BabaSec}

In this subsection, we investigate whether the WMAP data exhibits the type 
of large-scale symmetries predicted by small-universe models 
\cite{baba93,doc96}.

\subsubsection{The $S$-Statistic}

We perform our tests by computing the function 
$S(\hat{\bf n}_{i})$ defined by \cite{baba93,doc96}:
\beq{Smin_plane} 
     S(\hat{\bf n}_{i}) \equiv 
	\frac{1}{N_{pix}} \sum_{j=1}^{N_{pix}}  
	\left[ \frac{\delta T}{T}(\hat{\bf n}_j) - 
	\frac{\delta T}{T}(\hat{\bf n}_{ij}) \right]^{2},
\eeq 
where $N_{pix}$ is the number of pixels in the cleaned map and 
$\hat{\bf n}_{ij}$ denotes the reflection of $\hat{\bf n}_{j}$ in the plane 
whose normal is $\hat{\bf n}_{i}$, \ie, 
\beq{normal} 
	\hat{\bf n}_{ij} = 
	\hat{\bf n}_{j} - 2 (\hat{\bf n}_{i} \cdot \hat{\bf n}_{j}) 
	\hat{\bf n}_{i}.
\eeq
$S(\nh)$ is a measure of how much reflection 
symmetry there is in the mirror plane perpendicular to $\nh$. The 
more perfect the symmetry is, the smaller $S(\nh)$ will be. 

% When we calculate $S(\hat{\bf n}_{i})$ for all pixel positions 
% $\nh_i$ in the sky, we obtain a sky map that we refer to as an 
% $S$-map. 
We refer to the map of  $S(\hat{\bf n}_{i})$ over the whole sky as an  
$S$-map. This map is a useful visualization tool and gives 
intuitive understanding of how the $S$-statistic works.  
% $S$-maps are shown in Figure~\ref{fig1} and discussed in more detail 
% in \Sec{DataAnalysisSo}.
$S$-maps are discussed in more detail in \Sec{DataAnalysisSo}.

For a given sky map, we compute a test statistic $S_{\circ}$ defined as 
the minimum value of the $S$-map. As explained and illustrated in
\cite{doc96}, small-universe models with toroidal topology exhibit 
symmetries that give them on average smaller values of $S_{\circ}$ 
than infinite universes, which makes the $S$-statistic useful 
for constraining the cosmic topology.

\subsubsection{$S$-maps from the WMAP data}\label{DataAnalysisSo}

The $S$-map of our WMAP cleaned map from \cite{tegmark03}
is shown in Figure~\ref{fig1} (top right), and shows a striking 
pair of dark spots on the same preferred axis that we identified in
\Sec{AlignmentSec}, \eq{axisEq}, at ``two o'clock'' and its antipode.
Such an isolated minimum in the $S$-map is characteristic of
a $T^1$-model, where the Universe is a 3-torus with only one 
direction small relative to the cosmic horizon.
%\beq{positions}
%	(\ell,b) \approx (-85,48) % ~~  {\rm \&}  ~~ (94,-48),
%      [pix#,phi(l),theta(b)] = (42821,   85.9821,138.1412) for res64
%      [pix#,phi(l),theta(b)] = (  6325, 265.9821,  41.8587) for res64
%\eeq
%and its antipode
Figure~\ref{fig1} also shows the $S$-maps of the quadrupole 
(middle right) and the octopole (bottom right) components, and both are 
seen to independently show dark spots along this special axis.
% DO WE REALLY NEED TO SAY THIS, SINCE YOU'RE DISMISSING 
% THIS ENTIRE THING AS STATISTICALLLY INSIGNIFICANT ANYWAY 
% IN THE NEXT SECTION?
% Although these maps have regions of minima that are coincident 
% with the direction of  the two ``dark spots"  cited above, their 
% $S_{\circ}$ values lies in a plane perpendicular to the direction 
% of ``smallness" of the cleaned map. This happens because the 
% lobes seem to have a highly symmetric distribution in both cases. 
%%
In contrast, the $S$-maps of the hexadecapole and higher multipoles 
do not have this property.

\begin{figure}[tb]
%\preskip
\center{{\epsfxsize=9.0cm\epsffile{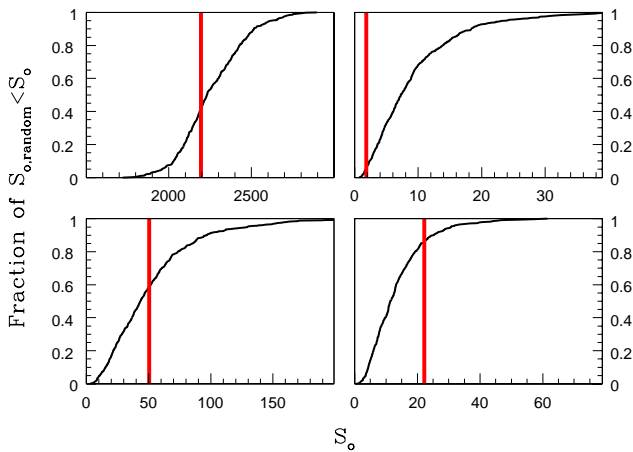}}}
\vskip-3.2cm
\caption{\label{fig2}\footnotesize%
	Cumulative histograms of $S_{\circ}$. {\it Top-left}: $S$-test for
	the cleaned WMAP map. {\it Top-right}: $S$-test for octopole alone.
	{\it Bottom-left}: $S$-test for the sum of the quadrupole and octopole.
	{\it Bottom-right}: $S$-test for hexadecapole alone. 
	Curves show the fraction of 500 simulated maps that have $S_{\circ}$ 
	below the given value. A small Universe should give a small $S_{\circ}$-value, 
	but the observed value of $S_{\circ}^{WMAP}$ (vertical line) is seen to be 
	significantly smaller than expected in an infinite universe only for the 
	octopole case.
	}
\end{figure}
 
\bigskip

\begin{figure}[tb]
\preskip
{\epsfxsize=12.5cm\epsffile{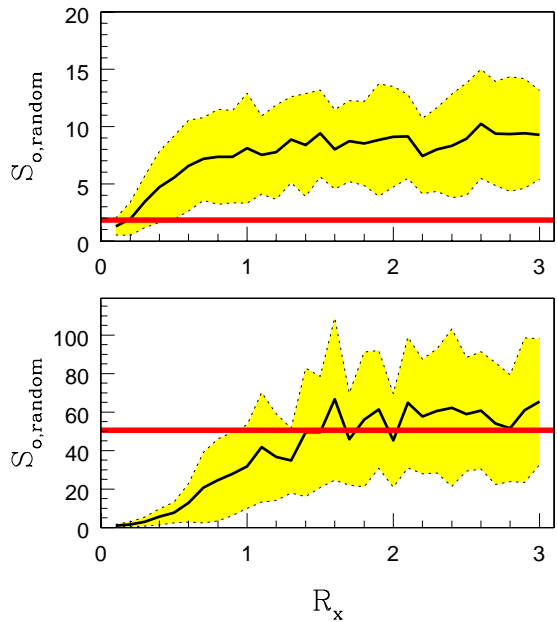}}
\vskip-4.3cm
\caption{\label{fig3}\footnotesize%
	Expected and observed $S_{\circ}$-values for the octopole map 
	alone (top) and for the sum of the quadrupole and octopole maps 
	(bottom). The solid line shows the mean of the Monte Carlo 
	simulations and the yellow (or grey in BW)  band shows the $1-\sigma$ 
	spread. 120 simulations per $R_x$-value were performed for $R_x\le 1$ 
	and 30 per $R_x$-value for  $R_x>1$. The horizontal lines represent 
	the observed values $S_{\circ}^{WMAP}$.
	}
\end{figure}

\subsubsection{What is the probability that an infinite Universe posseses such symmetry?}

To quantify the statistical significance of this symmetry axis, we produced 
500 all-sky Monte Carlo simulations with HEALpix resolution $nside$=16, 
with monopole and dipole removed, adopting $\ell_{max}$=128, and using 
the WMAP $\Lambda$CDM fiducial power spectra of \cite{spergel03}. We 
then use \eq{Smin_plane} to compute the $S$-map and $S_{\circ}$ for each 
of those 500 simulations.  
Figure~\ref{fig2} (top left) shows a cumulative histogram of $S_{\circ}$ for these 
500 simulations. For comparison, the vertical line corresponds to the value
of  $S_{\circ}^{WMAP}$ for the real WMAP data processed in the same way. 
The fact that $S_{\circ}^{WMAP}$ 
for the real data falls near the middle of the distribution means that although the 
dark spots in seen in Figure 1 (top right) may appear visually striking, symmetries 
at this level are  not at all unlikely to happen by chance in an infinite 
universe. In other words, the WMAP data is perfectly consistent with a 
standard infinite universe as far as our $S$-test is concerned. 
We obtain similar results when analyzing the octopole alone (top right), 
the hexadecapole alone (bottom right), various higher multipoles alone, 
as well as the sum of quadrupole and octopole  (bottom left).
For the case of the octopole (top right), however, we find the high degree 
of symmetry to be marginally significant, with an $S_{\circ}$-value as low 
as observed by WMAP occurring in less than 5\% of the simulations. If the 
octopole had all its hot and cold spots centered on a single plane, it would 
have perfect reflection symmetry about this plane, so this one in 20 anomaly
is simply another manifestation of the planar nature of the octopole 
discussed in \Sec{OctopoleSec}.
 
Note that there is nothing to be learned from applying the $S$-test 
to multipoles $\l=$0, 1 or 2 on their own, since (apart from pixelization 
effects) they all give $S_{\circ}=0$.
The monopole exhibits perfect symmetry around any axis, so its 
$S$-map is identically zero. The dipole has perfect reflection symmetry 
in all axes perpendicular to the dipole direction, so its $S$-map equals 
zero on a great circle. The quadrupole is a quadratic function given by 
a symmetric $3\times 3$ matrix, and therefore has perfect reflection 
symmetry about the three orthogonal eigenvectors of this matrix. This
is illustrated in Figure~\ref{fig1}, where the quadrupole $S$-map 
(middle right) exhibits six zeroes corresponding to the pair of hot spots, the 
pair of cold spots and the pair of saddle points in the quadrupole map 
(middle left).

\subsubsection{Making Toroidal Fake Skies}

Above we found that the $S$-statistic was consistent with an infinite universe.
Since a small universe could nonetheless explain all three of the anomalies
of \Sec{AnomalySec} in one fell swoop, let us investigate whether small universes
provide a better fit to the data than an infinite universe does or, conversely, 
whether the $S$-statistic rules out some interesting class of small universe 
models.

In a toroidal universe, only wave numbers that are harmonics of the cell size 
are allowed. Therefore, we have a discrete {\bf k} spectrum \cite{zeldovich73,fang87}
\beq{kcut} 
	{\bf k} = \frac{2 \pi}{R_H} 
	              \left(  \frac{p_x}{R_x}, \frac{p_y}{R_y},\frac{p_z}{R_z} \right),
 \eeq
where  $p_x,  p_y$ and $p_z $  are integers, and $R_x$,  $R_y$ and $R_z$ 
are the sizes of the cell. It was shown in \cite{doc96} that toroidal universes 
can be simulated by
\beq{new_SW} 
	\frac{\delta T}{T} (\theta,\phi) \propto
	\sum_{p_x, p_y, p_z}
	\left[ g_1 \cos(2 \pi \gamma) + g_2 \sin(2 \pi \gamma) \right] \alpha^{\frac{n-4}{4}},
\eeq
where  
\beq{gamma}
\gamma = \left( \frac{p_x}{R_x} x + \frac{p_y}{R_y} y + \frac{p_z}{R_z} z \right), 
\eeq
\beq{alpha}
\alpha \equiv {R_H|{\bf k}|\over 2\pi} = \left( \frac{p_x}{R_x} \right)^2 +
                         \left( \frac{p_y}{R_y} \right)^2 +
                         \left( \frac{p_z}{R_z} \right)^2, 
\eeq
$g_1$ and $g_2$ are two unit Gaussians random variables,  and $n$ 
is the spectral index\footnote{
	A more realistic Monte Carlo can be done by adding to each
	toroidal simulation a map that has only the ISW contribution. 
	A power suppression and enhanced symmetry at low multipoles would remain.
	}.
We generate our simulations with $n=1$ since this provides a good 
fit to the observed CMB power spectrum on the large angular scales 
that matter here (our tests below use only the information at the lowest 
multipoles). The symmetry patterns expected in $T^1$ universes have 
been shown to appear for broad range of $n$-values (actually, 
$n<$3) \cite{doc96}. Since we are focusing on $T^1$ universes, we wish to 
set $R_y=R_z=\infty$. In practice, we set $R_y=R_z=3$, since this is more 
numerically convenient and essentially indistinguishable in practice.
We add no noise to the simulations, since WMAP detector noise is completely 
negligible for $\l\le 20$.

\subsubsection{If we live in a $T^1$, can we constrain the cell size $R_x$?}

For the tests in this section, we create two band-pass filtered versions of
the foreground-cleaned WMAP map, retaining only $\l=3$ and only $2\le l\le 3$, 
respectively. We then degrade the resolution of these two maps to HEALpix 
resolution $nside$=16.
After generating our simulated small universe maps as above, we band-pass 
filter and process them in the same way as the real data. Since 
we wish to test only for symmetry properties, we normalize the two types of
Monte Carlo maps to have the same rms fluctuations as the two corresponding 
filtered WMAP maps.

Fixing a cell size, we constructed a simulated CMB map as described above and 
used \eq{Smin_plane} to  obtain an $S$-map from which we extracted its minimum
value $S_{\circ}$. Repeating this procedure 120 times, we obtain the 
probability distribution of $S_{\circ}$ for a given cell size. 
When we repeat
this procedure for different cell sizes we are able to construct Figure~\ref{fig3}. 
% pix size = sqrt(Area_all_sky/number_of_pixels)=sqrt(41252/3072)=3.6644deg
% FWHM~10deg

In Figure~\ref{fig3}, we show the probability distribution of $S_{\circ}$ obtained 
from Monte Carlo simulations for cell sizes $0.1<R_x<3$. Comparing this with 
the horizontal solid line (which represents the $S_{\circ}$-value extracted from 
the WMAP data, $S_{\circ}^{WMAP}$) shows what range of $R_x$-values 
WMAP favors.
The octopole alone (top panel) is seen to favor a small Universe with
$R_x<0.5$ at the $1-\sigma$ level, since it has near reflection symmetry 
about the preferred plane discussed in \Sec{OctopoleSec}. However, 
% both the quadrupole alone and 
the quadrupole-octopole combination (bottom panel) is seen to disfavor 
small universes, preferring $R_x\simgt 1$. 
We repeated the same test for all individual multipoles $\l=4,...,10$ and for 
the ranges $2-10$, $11-20$ and $2-20$, finding no significant preference 
for small universes. Combining the evidence of all these tests, we 
therefore conclude that the WMAP $S$-test does not favor the small 
universe hypothesis, preferring a cell size at least as large as our 
Hubble Volume, $R_x\simgt 1$.

\subsection{Searching for circles in the sky}\label{CircleSec}

The circles method \cite{css98} involves matching circles in the CMB 
sky.  The CMB Surface of Last Scattering (SLS) imaged by WMAP is a sphere 
with us at its center. The simple case of a toroidal Universe is mathematically 
equivalent to a perfectly periodic Universe, with copies of us on an infinite 
rectangular lattice. Since each copy of us is surrounded by its own spherical 
SLS, and the intersection of two spheres a circle, it follows that if the lattice
spacing is smaller than the diameter of the sphere, we can observe the 
same circular part of the SLS in opposite directions in the sky. In the case of
a  sufficiently small spacing, there will be several such matched circles, 
all concentric. 
As shown in \cite{css98}, this  result is general: {\it any} compact topology 
generates matched circles. For more complicated cases, these circles 
will typically no longer be opposite from one another as mirror images, 
but can differ in location, size, rotation and parity. 
%%
% To avoid misunderstandings, note that one does not see intrinsic circles in 
% the microwave sky; the two circle pairs do not have constant temperatures,
% but rather, the temperature varies identically around both circles.

\subsubsection{Real-world complications}

In the original discussion of this effect, \cite{css98} considered the idealized case 
where the entire CMB signal came from the SLS and would look the same 
from different vantage points. In practice, however, we need to consider three 
important departures from these assumptions:
\begin{enumerate}
	\item 
	The late integrated Sachs-Wolfe (LISW) effect is generated not on the 
	SLS but later on, by the gravitational potential that the CMB photons 
	traverse en route to us, so this contribution to the WMAP map will {\it not }
	match between paired circles.
	\item
	Whereas the CMB fluctuations generated by SLS density fluctuations 
	and gravitational potential fluctuations look the same from any vantage 
	point, those generated by the Doppler effect from velocity perturbations 
	do not. Rather, they have a dipole structure, which means that what looks 
	like a hot spot from one vantage point looks like a cold spot when viewed
	from the opposite direction.
	The Doppler contribution to CMB fluctuations therefore does not match
	between paired circles, and will in fact be strongly anti-correlated for small 
	matched circles (whose features we view from nearly opposite directions).
	\item
	The contribution to CMB maps from foreground contamination and detector 
	noise will not match between paired circles.
\end{enumerate}
 
To minimize the effect described in the third item, we based our analysis on the
foreground-cleaned WMAP map from \cite{tegmark03}, which minimizes the 
combined fluctuations of foregrounds and noise. Although residual foreground 
contamination may prevent us from finding a pair of small matched circles hiding 
in the Galactic plane, the above-mentioned axis that we are most interested in is
fortunately nowhere near this plane. 

We tackle the effects described in items {\it 1} and {\it 2} by filtering the WMAP 
map before performing the circle search. 
Since the LISW effect is important only for the very largest multipoles, we can
high-pass filter the WMAP map by zeroing all multipoles with $\l<8$, therefore 
removing the contributions from {\it 1}. 
The Doppler effect, on the other hand, is dominant only in the troughs between 
the acoustic peaks (it is mainly this effect that prevents the CMB power spectrum 
from dropping all the way down to zero between the peaks). Therefore, we can 
remove the contributions from {\it 2} by zeroing all multipoles with $\l>300$ as well. 
If we were to detect a suspected matched circle pair, a powerful subsequent 
consistency test would be whether an $\l=301-400$ band-pass filtered map 
(probing roughly the first acoustic trough) displays anti-correlation between the 
two circles.

The issue of how to best deal with the above-mentioned real-world complications 
for circle matching is an interesting and challenging data analysis problem, worthy 
of a future paper in its own right. Our analysis here should simply be viewed as a 
first attack on the problem, and the forthcoming analysis by \cite{spergel03b}, 
in preparation, will undoubtedly improve it.
One challenge is that band-pass filtering is spatially non-local, potentially reducing 
detectability by smearing out matched circles and by superimposing signal from
neighboring sky regions on the circles.
A second challenge is how to optimally deal with the fact that the WMAP map is 
pixelized. This means that there generally will not be a pixel centered exactly at 
the desired reflected position $\nh_{ij}$ in \eq{Smin_circle} below --- in this paper, 
we simply use the closest pixel. This pixelization problem is exacerbated by a 
desire to perform the circle search at as low resolution as possible to avoid the 
computations becoming prohibitively slow.

\subsubsection{The search} 

We limit our search to the simplest class of small-universe models, the ones 
where space is flat and toroidal. For this simple case, the matched circles come in 
diametrically opposed pairs, specified by three parameters: the position, given by ($l,b$) 
for the center of a circle, and its angular radius $\alpha$. A much more ambitious 
six-parameter search corresponding to the general case will be presented by 
Spergel {\etal} \cite{spergel03b}. 

To search for matched circles, we define a family of difference maps. For a given 
axis $\nh_i$, the map is defined through
\beq{Smin_circle} 
	D(\nh_i)\equiv \frac{\delta T}{T}(\nh_j) - \frac{\delta T}{T}(\nh_{ij}),
      %D(\nh_i)\equiv \left[\frac{\delta T}{T}(\nh_j) - \frac{\delta T}{T}(\nh_{ij})\right]^2,
\eeq 
where $\nh_{ij}$ denotes the reflection of $\nh_j$ in the plane whose normal 
is $\nh_i$ (\ie, $\nh_{ij}$ is given by \eq{normal}).
In other words, the map $D$ is just the original WMAP map minus its reflection 
about the plane normal to $\nh_i$. Note that \eq{Smin_circle} and \eq{Smin_plane}
are closely related: $S_0$ is simply the smallest average of a squared 
$D$-map.
In the ideal case of a perfectly matched circle pair, it would manifest itself as a
pair of $D=0$ circles in a $D$-map that otherwise fluctuates with about twice the 
variance of the original WMAP map. 
Note that in addition, there will always be a great ($\alpha=90^\circ$) circle with $D=0$ 
lying in the reflection plane: here 
the WMAP map is of course equal to its reflection, so the $D$-map will 
automatically vanish.

\begin{figure}[tb]
\center{{\epsfxsize=9.0cm\epsffile{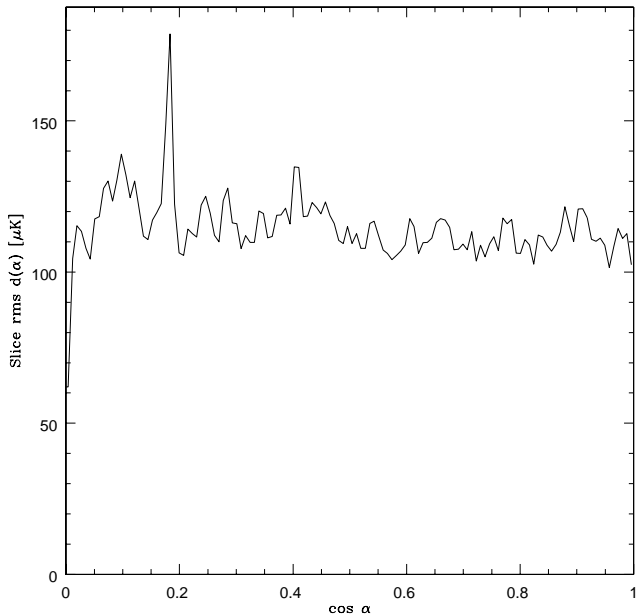}}} 
\vskip-0.5cm
\caption{\label{fig4}\footnotesize%
	An example of the curve $d(\alpha;\nh_i)$ that we used to search for matched
	circles. This case is for the reflection axis $\nh_i$ corresponding to 
	($l,b$)= $(-110^\circ,60^\circ)$. A pair of perfectly matched circles 
	of angular radius $\alpha$ would give $d(\alpha)=0$. The drop towards 
	zero on the left hand side is caused by a great ($\alpha=90^\circ$) circle 
	being its own reflection, whereas the high values around $\alpha \sim 15^\circ$ 
	are caused by residual foreground contamination.}
\end{figure}

To distill out this matched circle information, we replace each $D$-map by 
a single curve $d(\alpha;\nh_i)$ giving its rms along circles making a constant 
angle $\alpha$ with the reflection axis, \ie, with 
	$\nh_i\cdot\nh_j=\cos\alpha$.
An example of such a curve (for a fixed $\nh_i$) is shown in Figure~\ref{fig4}.
In practice, the HEALpix pixelization subdivides the sky into $12 n^2$ pixels, 
where the $n$ is a power of two determining the resolution. We compute 
$d(\alpha)$ in $n$ angular bins equispaced in $\cos\alpha$, which corresponds 
to a bin width about 0.87 times the pixel side $(4\pi/12)^{1/2}/n$ for $\alpha=30^\circ$. 
The advantage of this binning scheme is that each bin gets contributions from 
about the same number of pixels, $12n$.
We performed our analysis with $n=256$, which took a couple of weeks on a 
2 GHz linux workstation. For a given potential circle center $\nh_i$, the curve 
$d(\alpha;\nh_i)$ will thus oscillate randomly as a function of $\alpha$ with 
roughly constant variance, as shown in Figure~\ref{fig4}. 
Apart from pixelization effects, this curve will drop to zero at any $\alpha$ 
corresponding to a perfectly matched circle. Figure~\ref{fig4} shows such a 
deep minimum only for $\alpha=90^\circ$, the above-mentioned case of the 
$90^\circ$ circle which of course equals its own reflection. 

\begin{figure*}[tb]
\center{{\hglue0.8cm\epsfxsize=17.0cm\epsffile{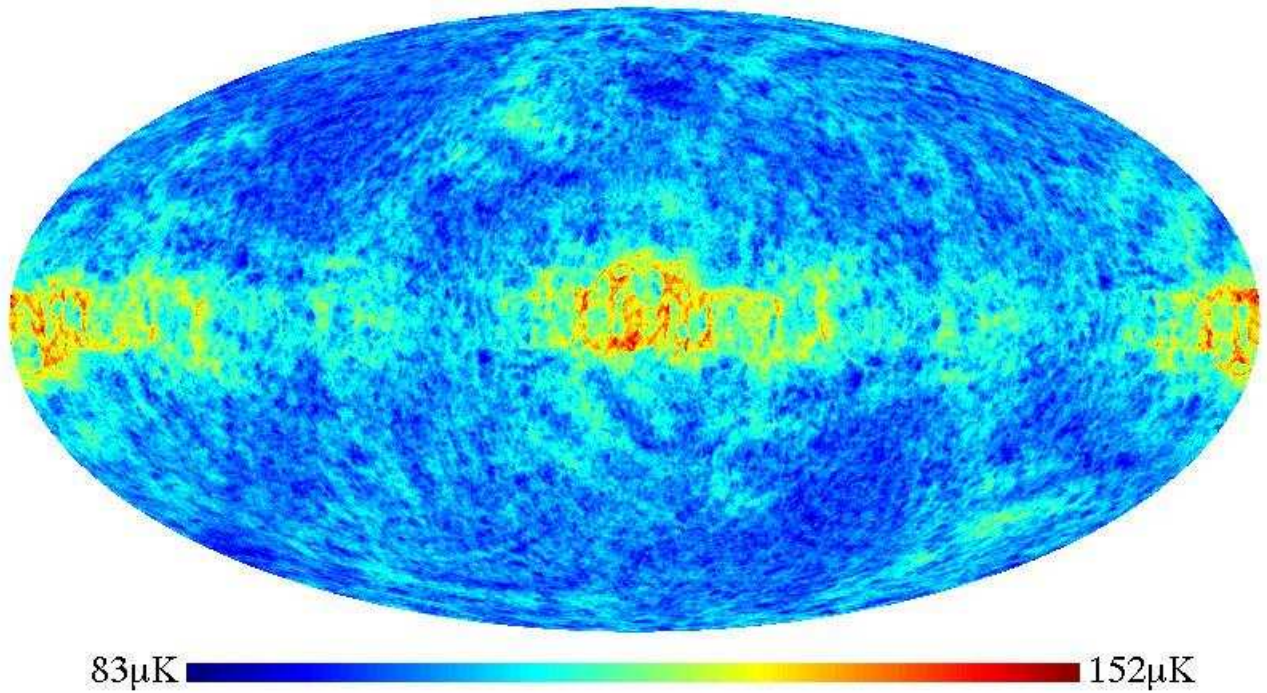}}}
%\vskip-2.8cm
\caption{\label{fig5}\footnotesize%
	Result of search for matched circles of radius $0^\circ<\alpha<15^\circ$.
	A perfectly matched circle would show up as a pixel with zero temperature 
	at the position of the circle center, whereas the map above shows no pixels
         below 83$\mu$K. Note that this map is parity-symmetric, \protect\ie, the temperature
        at $\nh$ equals that at $-\nh$, although this symmetry is obscured by the Aitoff
        projection used.
}
\end{figure*}
 
To analyze the results of our search, we compute a summary map $\dmin$,
defined as
\beq{dminDefEq} 
	\dmin(\nh_i) \equiv \min_{\alpha_0<\alpha<\alpha_1} d(\alpha;\nh_i).
\eeq
If matched circles are present, the summary map should show evidence f
or these circles in the radius range from $\alpha_0$ to $<\alpha_1$, and each 
pixel in this map should correspond to the evidence for circles centered on 
that pixel.
 
Figure~\ref{fig5} shows the summary map for circle radii in the range 
$0^\circ<\alpha<15^\circ$. The color scale is seen to range over about a factor 
of two in $d$-value (the units are $\mu$K). A matched circle would correspond 
to a pair of diametrically opposite pixels with values near zero.
Not only are no such pixels seen, but the map shows that the region around
($l,b$)= $(-110^\circ,60^\circ)$ that we are particularly interested in given 
the quadrupole-octopole anomalies is in no way unusual compared to the 
rest of the map.
The larger values seen near the equator of the map are caused by residual 
foreground contamination, which from the definition of the $\dmin$ maps 
propagates about one circle radius from the Galactic plane. Whereas this 
foreground contamination can mask true matched circles centered in this 
region, it can clearly never create false positives, since it increases rather 
than decreases the $\dmin$-value.

We computed analogous maps for four other ranges of circle sizes $\alpha$,
	$15^\circ -30^\circ$,
	$30^\circ -45^\circ$,
	$45^\circ -65^\circ$,
	$65^\circ -85^\circ$,
again finding no evidence of matched circles. For completeness, we also 
computed a final map for the $\alpha$-range 
	$85^\circ -90^\circ$ 
obtaining roughly zero, since all $90^\circ$ circles are matched with 
themselves.

In conclusion, we find no evidence for the matched circles that are predicted 
for a simple toroidal universe. 

\section{Conclusions}\label{ConclusionSec}

We have investigated anomalies reported in the Cosmic Microwave Background 
maps from the WMAP satellite on very large angular scales. There are three 
independent anomalies involving the quadrupole and octopole:
\begin{enumerate}
	\item 
	The cosmic quadrupole on its own is anomalous at the 1-in-20 level by 
	being low (the cut-sky quadrupole measured by the WMAP team is more 
	strikingly low, apparently due to a coincidence in the orientation of our 
	Galaxy of no cosmological significance).
	\item 
	The cosmic octopole on its own is anomalous at the 1-in-20 level by 
	being very planar.
	\item 
	The alignment between the quadrupole and octopole is anomalous at 
	the 1-in-60 level.
\end{enumerate}
Although the {\it a priori} chance of all three occurring is 1 in 24000, the multitude 
of alternative anomalies one could have looked for dilutes the significance of 
such {\it a posteriori} statistics.

However, in the context of 
constraining cosmic topology, these three effects are not merely 
three arbitrary ones among many. Indeed, 
the simplest small Universe model where the Universe has toroidal topology 
with one small dimension of order half the horizon scale, in the direction towards 
Virgo, could explain {\it 1, 2} and {\it 3}. In order to test this hypothesis, we applied
the $S$-statistic and the circle test on the WMAP data, ruling out this model. 
In other words, we have ruled out the ``plain bagel'' small universe model.

Our results also rule out other models that predict back-to-back matched circles.
However, they do not rule out the recently proposed 
dodecahedron model of \cite{Luminet03}: although this model predicts six pairs of
diametrically opposed circles of radius about $35^\circ$, the circles have a 
$36^\circ$ twist relative to their twin images, thereby eluding our search method.
After the original version of this paper had been submitted, a 
more thorough analysis by Cornish and collaborators \cite{Cornish03}
confirmed our findings and improved them to rule out 
this and other twisted back-to-back models as well.

A maximally ambitious six-parameter ``everything bagel'' circle search, corresponding to 
the general case 
of arbitrary topologies, is currently being carried out by Spergel and collaborators,
and will be presented in a forthcoming paper \cite{spergel03b}. This should provide 
decisive evidence either for or against the small universe hypothesis. 
If this circle search confirms our finding that small universes cannot explain the 
anomalies, we will be forced to either dismiss the anomalies as a statistical fluke 
or to search for explanations elsewhere, such as modified inflation models
\cite{bridle03,uzan03a,contaldi03,cline03,feng03,efstathiou03}.
Even the fluke hypothesis might ultimately be testable, since it may be possible 
to improve the signal-to-noise of the large scale power spectrum beyond the 
WMAP cosmic variance limit by employing cluster polarization \cite{kamion97,seto00} 
or weak gravitational lensing \cite{kesden03} techniques.

\bigskip
\bigskip

We thank the the WMAP team for producing such a superb data set and for 
promptly making it public via the Legacy Archive for Microwave Background 
Data Analysis (LAMBDA) -- the WMAP data are available from 
	\url{http://lambda.gsfc.nasa.gov}.
Support for LAMBDA is provided by the NASA Office of Space Science.
We thank Krzysztof G\'orski and collaborators for creating the HEALPix package 
\cite{healpix1,healpix2}, which we used both for harmonic transforms and map 
plotting.
Thanks to James Bjorken for stimulating discussions.
This work was supported by NASA grant NAG5-11099. 
MT is supported by NSF grant AST-0205981 and a Cottrell Scholarship 
from Research Corporation.
MT \& MZ are supported by David and Lucile Packard Foundation fellowships.
MZ is supported by NSF and 
AJSH is supported by NSF grant AST-0205981 and NASA grant NAG5-10763.

\clearpage

\appendix

\section{Rotation}

In this appendix, we review the rotation properties of spherical harmonics
that were used in \Sec{AlignmentSec} and \Sec{OctopoleSec}.
This material is well-known. %None of this is material is new. 
However, since we found it rather time-consuming to assemble these 
explicit results from the literature (which contains a variety of 
notational conventions and is mainly geared towards the general 
quantum case where functions are complex rather than real), 
we summarize the results here for the benefit of the reader interested in
performing similar calculations. Fortran77 software implementing this is 
available from the authors upon request.

\subsection{Real-valued spherical harmonics}

When a function $\psi(\nh)$ is real-valued, the corresponding 
spherical harmonic coefficients obey $a_{\l-m}= (-1)^m\alm^*$.
The need to work with complex numbers is conveniently eliminated 
by working with real-valued spherical harmonics, which are obtained 
from the standard spherical harmonics by replacing $e^{im\phi}$
in their definition by $\sqrt 2\sin m\phi$, $1$, $\sqrt 2\cos m\phi$ for 
$m<0$, $m=0$, $m>0$, respectively. In other words, the vector 
of $2\l+1$ complex numbers $\alm$ ($m=-\l,...,\l$) specifying the $\lth$ 
harmonic is replaced by a vector of $2\l+1$ real numbers $\atlm$ given 
by $\sqrt{2}\>{\rm Im}\>\alm$, $a_{00}$, $\sqrt{2}\>{\rm Re}\>\alm$.
This mapping from $\alm$ to $\atlm$ corresponds to multiplication by 
a unitary matrix $\U$; for example, the $\l=3$ case corresponds to
\beq{U3Eq}
\U=\left(
\begin{matrix}
     {\frac{-i}{{\sqrt{2}}}} & 0 & 0 & 0 & 0 & 0 & {\frac{-i}{{\sqrt{2}}}} \cr 
     0 & {\frac{i}{{\sqrt{2}}}} & 0 & 0 & 0 & {\frac{-i}{{\sqrt{2}}}}  & 0 \cr 
     0 & 0 & {\frac{-i}{{\sqrt{2}}}} & 0 & {\frac{-i}{{\sqrt{2}}}} & 0 & 0 \cr 
     0 & 0 & 0 & 1 & 0 & 0 & 0 \cr 0 & 0 & -{\frac{1}{{\sqrt{2}}}} 
		& 0 & {\frac{1} {{\sqrt{2}}}} & 0 & 0 \cr 
     0 & {\frac{1}{{\sqrt{2}}}} & 0 & 0 & 0 & {\frac{1}{{\sqrt{2}}}} & 0 \cr 
     -{\frac{1} {{\sqrt{2}}}} & 0 & 0 & 0 & 0 & 0 & {\frac{1}{{\sqrt{2}}}} \cr   
\end{matrix}
\right).
\eeq
By unitarity, $\sum_\l |\alm|^2 = \sum_\l \atlm^2$, and it follows that
the desired right-hand-side of \eq{Leq} is simply $\sum_\l m^2\atlm^2$.
The real-valued spherical harmonic coefficients $\atlm$ observed 
by WMAP are listed in Table~\ref{tab3} for $\l=2,3$.

\subsection{Rotations}

An arbitrary three-dimensional rotation is specified by three Euler 
angles $(\phi,\theta,\alpha)$, defining a rotation by $\alpha$ around 
the $z$-axis followed by a rotation by $\theta$ around the $y$-axis
followed by a rotation by $\phi$ around the $z$-axis.
The $z$-rotation by $\phi$ has a trivial effect on the spherical harmonic 
coefficients, simply multiplying $\alm$ by a phase factor $e^{i m\phi}$, 
so the Euler angle $\phi$ will not affect the absolute value $|\alm|$ 
and hence the quantity computed in \eq{Leq}.
The axis $\nh$ mentioned in \Sec{AlignmentSec} thus does not involve 
$\phi$ and is defined by 
	$\nh=(\sin\theta\cos\alpha,\sin\theta\sin\alpha,\cos\theta)$.
The $y$-rotation multiplies the vector of $(2\l+1)$ $\alm$-coefficients 
for the $\lth$ multipole by the Wigner matrix $\D^{\l}(\theta)$ given by
\cite{edmonds57,hamilton93,risbo96}
% According to http://wwwinfo.cern.ch/asdoc/shortwrupsdir/u501/top.html
\beqa{dzEq}
&\D&^{\l}_{mm'}(\theta) = 
	\sum_{k={\rm max}\{0,m+n\}}^{{\rm min}\{\l+m,\l+n\}}
	 \left(\cos{\theta\over 2}\right)^{2k-m-n} 
	 \left(\sin{\theta\over 2}\right)^{2\l+m+n-2k}\nonumber\\
	&\times& (-1)^{k+\l+m} {\sqrt{(\l+m)!(\l-m)!(\l+n)!(\l-n)!} 
	\over k!(\l+m-k)!(\l+n-k)!(k-m-n)!}.
\eeqa
Transforming to our real-valued spherical harmonic basis, this corresponds to 
the rotation matrix $\R_y(\theta)\equiv \U\D^{\l}(\theta)\U^\dagger$ given by 
\beq{QuadrupoleRotationEq}
\R_y(\theta) = 
\left(
\begin{matrix} 
     \cos \theta & -\sin \theta & 0 & 0 & 0 \cr
     \sin \theta & \cos \theta & 0 & 0 & 0 \cr 
     0 & 0 & {\frac{1 + 3\,\cos 2\theta}{4}} & {{\frac {\sqrt{3}}\,\sin 2\theta}{2}} & \frac{1 - \cos 2\theta}{4/\sqrt{3}} \cr 
     0 & 0 & {\frac{-{\sqrt{3}}\,\sin 2\theta }{2}} & \cos 2\theta & {\frac{\sin 2\theta}{2}} \cr 
     0 & 0 & \frac{1 - \cos 2\theta}{4/\sqrt{3}} & {\frac{-\sin 2\theta}{2}} & {\frac{3 + \cos 2\theta}{4}} \cr  
\end{matrix}    
\right)
\eeq
for the quadrupole case, and the corresponding matrix for rotations around 
the $z$-axis is
\beq{QuadrupolezRotationEq}
\R_z(\varphi) = 
\left(
\begin{matrix}  
  \cos 2\varphi & 0 & 0 & 0 & \sin 2\varphi \cr 
  0 & \cos\varphi & 0 & \sin\varphi & 0 \cr 
  0 & 0 & 1 & 0 & 0 \cr
  0 & -\sin\varphi & 0 & \cos\varphi & 0 \cr 
  -\sin 2\varphi & 0 & 0 & 0 & \cos 2\varphi \cr  
\end{matrix} 
\right).
\eeq
For the octopole case, \eq{QuadrupoleRotationEq} is replaced by the 
block-diagonal matrix
\beq{OctopoleRotationEq}
\R_y(\theta) = 
\left(
\begin{matrix} 
          \A	&\Zero\cr
          \Zero	&\B
\end{matrix} 	  
\right),
\eeq
where
\beq{Aeq}
\A = 
\left(
\begin{matrix} 
      {\frac{5 + 3\,\cos 2\theta}{8}} 
      & {\frac{- {\sqrt{{\frac{3}{2}}}}\,\sin 2\theta  }{2}} 
      & {\frac{1 - \cos 2\theta}{8/\sqrt{15}}} 
      \cr 
      {\frac{{\sqrt{{\frac{3}{2}}}}\,\sin 2\theta}{2}} 
      & \cos 2\theta 
      & {\frac{-{\sqrt{{\frac{5}{2}}}}\,\sin 2\theta }{2}}
      \cr 
      {\frac{1 - \cos 2\theta}{8/\sqrt{15}}}  
      & {\frac{{\sqrt{{\frac{5}{2}}}}\,\sin 2\theta}{2}} 
      & {\frac{3 + 5\,\cos 2\theta}{8}}
      \cr 
\end{matrix} 
\right)
\eeq
and
\beq{Beq}
\B = 
\left(
\begin{matrix} 
    {\frac{3\,\cos \theta + 5\,\cos 3\theta}{8}} 
    & {\frac{\sin\theta + 5\sin 3\theta}{16/\sqrt{6}}} 
    & {\frac{\cos\theta - \cos 3\theta}{8/\sqrt{15}}} 
    & {\frac{3\sin\theta - \sin 3\theta}{16/\sqrt{10}}} 
    \cr 
    {\frac{-\sin\theta - 5\sin 3\theta}{16/\sqrt{6}}} 
    & {\frac{\cos \theta + 15\,\cos 3\theta}{16}} 
    & {\frac{-\sin\theta + 3\sin 3\theta}{16/\sqrt{10}}} 
    & {\frac{\cos\theta - \cos 3\theta}{16/\sqrt{15}}} 
    \cr 
    {\frac{\cos\theta - \cos 3\theta}{8/\sqrt{15}}} 
    & {\frac{\sin\theta - 3\sin 3\theta}{16/\sqrt{10}}} 
    & {\frac{5\,\cos\theta + 3\,\cos 3\theta}{8}} 
    & {\frac{5\sin \theta + \sin 3\theta}{16/\sqrt{6}}} 
    \cr 
    {\frac{-3\sin\theta + \sin 3\theta}{16/\sqrt{10}}} 
    & {\frac{\cos\theta - \cos 3\theta}{16/\sqrt{15}}} 
    & {\frac{-5\sin\theta - \sin 3\theta}{16/\sqrt{6}}} 
    & {\frac{15\,\cos \theta +\cos 3\theta}{16}} 
      \cr  
\end{matrix} 
\right).
\eeq

The generalization of \eq{QuadrupolezRotationEq} to arbitrary 
multipoles is obvious.

\end{document}